\documentclass[aps,prl,twocolumn,groupedaddress]{revtex4}
\usepackage{amsmath}
\usepackage{amssymb}
\usepackage{pstricks}
\usepackage{graphics}
\usepackage{graphicx}
\providecommand{\av}[1]{\left\langle #1\right\rangle}

\providecommand{\ket}[1]{|#1\rangle}
\providecommand{\bra}[1]{\langle#1|}
\providecommand{\kebra}[2]{\ket{#1}\bra{#2}}

\providecommand{\tr}[1]{\textrm{tr}\left\{#1\right\}}
\providecommand{\tra}[1]{\textrm{tr}_a\left\{#1\right\}}
\begin{document}
\title{Oscillator state reconstruction via tunable qubit coupling in Markovian environments}
\author{Tommaso Tufarelli$^1$,  M. S. Kim$^2$, Sougato Bose$^1$}
\affiliation{$^1$Department of Physics and Astronomy, University College London, Gower Street, London WC1E 6BT, UK\\
$^2$QOLS, Blackett Laboratory, Imperial College London, SW7 2BW, UK}
\begin{abstract}
We show that a parametrically coupled qubit can be used to fully reconstruct the quantum state of a harmonic oscillator, even when both systems are subject to decoherence. By controlling the coupling strength of the qubit over time, the characteristic function of the oscillator at any phase space point can be directly measured by combining the expectation values of two Pauli operators. The effect of decoherence can be filtered out from the measured data, provided a sufficient number of experimental runs is performed. In situations where full state reconstruction is not practical or not necessary, the method can still be used to estimate low order moments of the mechanical quadratures. We also show that in the same framework it is possible to prepare superposition states of the oscillator. The model is very general but particularly appropriate for nanomechanical systems.
\end{abstract}
\maketitle
\section{Introduction}
With an increasing number of experiments bringing mesoscopic oscillators to the quantum regime, it has become essential to develop efficient readout methods that are able to verify quantum effects in such systems. The ultimate goal is the full reconstruction of the system's density matrix. Most of the current schemes to measure nanomechanical motion consist of weak continuous measurements of the oscillator's position (see e.g. \cite{meas-review}). Even though these techniques are useful for the monitoring of ground state fluctuations, they are not easily applicable to arbitrary nonstationary states, where the relation between the quantum state and the measured data is extremely complicated. Recently a method that allows full reconstruction of an arbitrary state of a mechanical oscillator has been proposed in the context of optomechanics \cite{pulsed}, however it relies on tomography techniques that involve transformations of the measured data. In both cases it is challenging to identify an unknown quantum state with high accuracy.

The idea that a two-level system (or qubit) could be used as a probe to reconstruct the quantum state of a continuous-variable system has been initially proposed in the field of cavity-QED, due to the impossibility to directly measure the intracavity field \cite{ms-kim,similar-cqed}. Similar ideas have subsequently been introduced in the field of circuit-QED \cite{circuit-qed}, ion traps \cite{ion-trap} and nanomechanical oscillators \cite{Meystre}. In all these works, information about the initial state of the oscillator is inferred from the transition probabilities of the qubit, after the two systems have interacted. Typically, the value of the characteristic function \cite{barnett} of the oscillator at several phase space points can be directly extracted from the qubit statistics. 

\textit{In this paper we report a further advance in this direction}. Firstly, we assume to control the parametric coupling between the qubit and oscillator as a function of time, a very interesting possibility opened by some recent experimental and theoretical works \cite{LaHaye,flux-oscill,gambetta-houck,srinivasan}. With this choice, any point of the oscillator's phase space can be reached by appropriately tailoring the time dependence of the coupling. The corresponding value of the characteristic function is then measured by combining the expectation values of two Pauli operators of the qubit. Differently from methods involving resonant interactions \cite{ms-kim,Meystre,circuit-qed}, no auxiliary degrees of freedom or additional displacement operations are required in our scheme. 

The second contribution we give to the field is the full inclusion of decoherence in the treatment, since to the best of our knowledge the effect of noise on such state reconstruction procedures has yet to be explored systematically. The inclusion of decoherence is essential especially when dealing with mesoscopic systems, where the coupling of \textit{both} the qubit and the oscillator to a thermal environment cannot be neglected. When the environment is described via the standard Markovian master equation, we obtain the remarkable result that the characteristic function can still be reconstructed in full detail. The price to pay is that a larger number of experimental runs will be needed to extract the same amount of information as compared to the ideal decoherence-free case. In practice, decoherence will limit the size of the phase space region where the characteristic function can be measured in a reasonable number of runs. When only the nearbies of the origin can be accessed, our method still gives useful information by providing low order moments of the mechanical quadratures.

As an addition we show, and examine, how well our system can be used to prepare superposition states of the oscillator in the presence of decoherence, and we provide a detailed analysis of the resulting characteristic function.

Having proposed a method to reconstruct the characteristic function, we remark that it constitutes a very useful description of a quantum state, even though it is sometimes overlooked in favor of the Wigner representation \cite{barnett}. As we emphasize in this paper, a number of properties such as purity, squeezing or the presence of coherent superpositions can be investigated from it as directly as from the Wigner function. Since any moment of the mechanical quadratures can be extracted from it in a straightforward way, the characteristic function is the ideal representation to invesigate statistical properties such as squeezing or non-Gaussianity \cite{Genoni}. Moreover, an efficient method to estimate nonclassicality directly from a finite collection of measured characteristic function values has been recently proposed \cite{mari}.
\section{The Model}
We consider a qubit with ground state $\ket{g}$, excited state $\ket e$, bare energy $\omega_q$ and tunneling energy $\delta$, coupled to a harmonic oscillator of frequency $\Omega$. The coupling strength $g(t)$ is for now a generic function of time. The Hamiltonian, in units of $\hbar=1$, is
\begin{equation}
H=\Omega a^\dagger a+\frac{\omega_q}{2}\sigma_z+\frac{\delta}{2}\sigma_x+g(t)\sigma_z(a+a^\dagger),
\end{equation}
where the Pauli matrices are $\sigma_x=\kebra{e}{g}+\kebra{g}{e},\sigma_y=-i(\kebra{e}{g}-\kebra{g}{e}),\sigma_z=\kebra{e}{e}-\kebra{g}{g}$, while $a$ is the annihilation operator for the harmonic oscillator. In a later section we will come back to some possible experimental realizations of such time-varying coupling. For the purposes of this paper, we will consider the parametric coupling regime $\delta=0$. In an interaction picture with respect to the free Hamiltonian, we have
\begin{equation}
H_I=g(t)\sigma_z\left(a e^{-i\Omega t}+a^\dagger e^{i\Omega t}\right).\label{int-picture}
\end{equation}
The open dynamics of the system is described by the master equation
\begin{equation}
\dot\rho=-i[H_I(t),\rho]+\mathcal L_m\rho+\mathcal L_q\rho+\mathcal L_p\rho,\label{master}
\end{equation}
where the non-unitary contributions to the dynamics are given by the Lindblad operators
\begin{align}
&\mathcal L_m\rho=\frac{\kappa}{2}(N_m+1)\mathcal D[a]\rho+\frac{\kappa}{2}N_m\mathcal D[a^\dagger]\rho,\label{lind-1}\\
&\mathcal L_q\rho=\frac{\Gamma_1}{2}(N_q+1)\mathcal D[\sigma_-]\rho+\frac{\Gamma_1}{2}N_q\mathcal D[\sigma_+]\rho,\\
&\mathcal L_p\rho=\frac{\Gamma_2}{2}\mathcal D[\sigma_z]\rho.\label{lind-3}
\end{align}
In the above, $\sigma_-=\kebra{g}{e}$, $\sigma_+=\kebra{e}{g}$, $\mathcal D[\hat A]\rho=2\hat A\rho\hat A^\dagger-\hat A^\dagger\hat A\rho-\rho\hat A^\dagger\hat A$, where $\hat A$ is a generic operator, $\kappa$ is the damping rate of the oscillator, $\Gamma_1$ and $\Gamma_2$ are respectively the damping and dephasing rates of the qubit, while $N_m$ ($N_q$) is the number of thermal excitations of the oscillator's (qubit's) environment.
\section{Characteristic function reconstruction}
To begin our state reconstruction protocol, we assume to be able to initialize the system in the separable state
\begin{equation}
\rho_\textrm{tot}(0)=\kebra{+}{+}\otimes\rho_0,\label{initial}
\end{equation}
where $\rho_0$ is the state of the harmonic oscillator that we want to reconstruct, and $\ket +=\tfrac{1}{\sqrt2}(\ket g+\ket e)$ \cite{pure-qubit}. After an interaction time $t$ the system evolves to a state $\rho_\textrm{tot}(t)$ according to (\ref{master}). At this point, we measure either the observable $\sigma_x$ or $\sigma_y$, which completes a single ``run'' of the experiment. A sufficient number of runs will have to be performed in order to estimate the average values $\av{\sigma_j(t)}=\tr{\rho_\textrm{tot}(t)\sigma_j}$ to the desired accuracy. By integrating Eq. (\ref{master}) as shown in the Appendix, we get our main result
\begin{equation}
\av{\sigma_x(t)}+i\av{\sigma_y(t)}=\chi\left(\xi(g,t)\right)e^{-f(g,t)},\label{main-result}
\end{equation}
where $\chi(\beta)=\textrm{tr}\left\{\rho_0 D(\beta)\right\}$ is the characteristic function, $D(\beta)=e^{\beta a-\beta^*a^\dagger}$ being the displacement operator \cite{barnett}. The notation $(g,t)$ indicates that $\xi$ and $f$ are dependent on the specific realization of the time dependent coupling $g(t)$, i.e. they are \textit{functionals}. Their explicit form is
\begin{align}
&\xi(g,t)=2i\int_0^tdsg(s)e^{i\Omega s-\frac{\kappa}{2}s}\label{xi_t}\\
&f(g,t)=\gamma t+\Delta(1-e^{-\kappa t})\left|\mu(g,t)\right|^2+\kappa\Delta\int_0^tds\left|\mu(g,s)\right|^2\label{f_t}\\
&\mu(g,t)=\frac{2i}{\sinh{\frac{\kappa}{2}t}}\int_0^tdsg(s)e^{i\Omega s}\sinh{\frac{\kappa}{2}s}\label{mu_t},
\end{align}
with $\gamma=\Gamma_1(N_q+1/2)+2\Gamma_2$ and $\Delta=N_m+1/2$. To reconstruct the characteristic function of the state $\rho_0$, one needs to ``invert`` Eq. (\ref{xi_t}), i.e. establish a mapping that associates any phase space point $\beta$ to an appropriate coupling and interaction time $(g_{\beta},t_{\beta})$, such that $\xi(g_{\beta},t_{\beta})=\beta$. One explicit example of such mapping is given in the next section.

As shown in Eq. (\ref{main-result}), the quantity that we can directly measure is given by the characteristic function evaluated at the phase space point $\xi(g,t)$ and ``damped'' by a factor $e^{-f(g,t)}$ due to decoherence. Remarkably, this is still a valid representation of the state $\rho_0$, since the functional $f(g,t)$ does not depend on the state of the oscillator and it is fully known in our theory once the coupling $g(t)$ is assigned \cite{s-parametrized}. The actual value of the characteristic function can thus be recovered \textit{even in the presence of finite decoherence}. This however requires that we measure the expectation values in (\ref{main-result}) with an accuracy greater than $e^{-f}$. Assuming that the relative error on $\av{\sigma_j}$ scales as $M^{-1/2}$, where $M$ is the number of experimental runs, $M\gg e^{2f}$ is needed to measure the value of the characteristic function at the point $\xi$ with sufficient accuracy. As we will see, $f$ tends to increase as $|\xi|$ is increased, meaning that decoherence imposes practical limits to our ``reach'' in phase space.
\section{Harmonic coupling}
\begin{figure}[t!]
\includegraphics[width=0.24\textwidth]{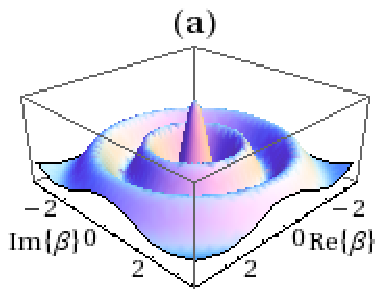}\includegraphics[width=0.24\textwidth]{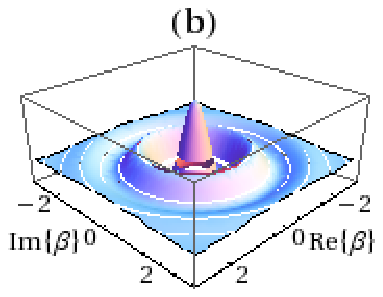}\\
\begin{center}
\includegraphics[width=0.24\textwidth]{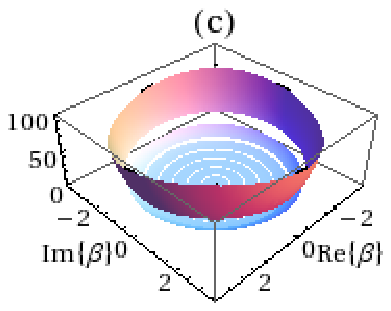}\end{center}
\caption{Plot (a): Characteristic function of the Fock state $\ket{5}$. Plot (b): Phase-space representation of the same state as directly measured from the qubit, according to Eq.(\ref{main-result}). Plot (c) : Value of $e^{2f}$ as a function of the phase space point $\beta$, giving a lower bound to the number $M$ of required experimental runs. Plots (b) and (c) are obtained with the damping exponent of Eq. (\ref{damping-approx}), using the parameters $r_0=0,r\leq0.5, \Omega=2\pi\times100$MHz, $\kappa=2\pi\times50$kHz,$ \gamma=\kappa\Delta=2\pi\times1$MHz (these parameters are derived from \cite{flux-oscill} in the implementations section of this paper). We can notice how the damping factor $e^{-f}$ affects more significantly the characteristic function away from the origin, resulting in a larger number of required experimental runs. Discontinuities in plots (b) and (c) are due to the discrete nature of the number of periods $n$ in Eq. (\ref{damping-approx}). As emphasized in the text, the characteristic function of plot (a) can be fully recovered from the measured data of plot (b), provided $M\gg e^{2f}$ runs are performed at each point $\beta$.\label{fock-plots}}
\end{figure}

An essential step of the reconstruction protocol is the inversion of Eq. (\ref{xi_t}). The realization of such inversion is clearly not unique and the specific form of $g(t)$ and the choice of the interaction time can be optimized to best suit the available experimental apparatus. To give a concrete example, we will consider a coupling constant that oscillates harmonically at the mechanical frequency:
\begin{equation}
g_{r,\phi}(t)= \frac{\Omega}{2\pi}e^{\frac{\kappa}{2}t}\left[r_0+r\sin\left(\phi-\Omega t\right)\right]\label{g_t},
\end{equation}
where the exponential factor $e^{\kappa t/2}$ is included to simplify calculations, $r>0$, while $r_0$ is a constant that might be needed to keep $g_{r,\phi}(t)$ inside the experimentally allowed range (for example, the constraint $g_{r,\phi}(t)>0$ would impose $r_0>r$). For simplicity, we restrict the possible interaction times to integer multiples of the mechanical period. This choice is not mandatory, however it considerably simplifies the task of inverting Eq. (\ref{xi_t}), allowing us to keep the treatment analytical and compact. Evaluating (\ref{xi_t}) for $t_n=n2\pi/\Omega$, with $n$ integer, gives
\begin{equation}
\xi\left(g_{r,\phi},t_n\right)=nre^{i\phi}.\label{space-point}
\end{equation}
We see that the amplitude and phase of $\xi$ are now related in a very simple way \textit{to the amplitude and phase of} $g(t)$. Had we chosen a generic interaction time, the value of $\xi$ would have depended in a more complicated way on the parameters $(r_0,r,\phi,t)$. For high frequency nanomechanical oscillators it is often the case that $g_\textrm{max}\ll\Omega$ \cite{LaHaye,flux-oscill}, thus the maximum value $r_\textrm{max}$ of the parameter $r$ will be typically small. A number of mechanical periods $n>1$ could then be used to reach phase space points of modulus greater than $r_\textrm{max}$. A possible inversion of Eq. (\ref{xi_t}) is then realized by associating the desired value of $\xi$ to the corresponding parameters $(r,\phi,n)$, where $n$ is the integer verifying $(n-1)r_\textrm{max}\leq|\xi|<nr_\textrm{max}$, while $r=|\xi|/n$ and $\phi=\arg{\xi}$. In practice the maximum number of cycles $n$ and thus the modulus of $\xi$ will be limited due to decoherence, as we now show by explicit calculation. The damping exponent $f$ defined in (\ref{f_t}) is in this case a complicated function of $(n,r,r_0,\phi)$. It is possible to obtain a compact result, valid for high quality factor oscillators, by approximating $f$ to first order in $\kappa/\Omega$:
\begin{align}
&f\left(g_{r,\phi},t_n\right)\simeq \gamma t_n +\kappa\Delta t_n\left[2 \frac{r_0^2}{\pi^2}-r_0r \frac{\sin\phi+2n\pi\cos\phi}{2\pi^2}+\right.\nonumber\\
&\phantom{f\left[g_{x_0,y_0},t_n\right]}\left.+r^2\left(\frac{n^2}{3}+\cos\phi\frac{1-2n\pi\sin\phi}{4\pi^2}\right)\right].\label{damping-approx}
\end{align}
If the parameters $(r_0,r,\phi)$ are kept fixed, it can be shown that (\ref{damping-approx}) is a strictly increasing function of the number of periods $n$, confirming the intuition that decoherence tends to be more severe for regions of phase space away from the origin \cite{explain}. As an example, if $r_\textrm{max}\simeq0.5, \kappa\Delta\simeq0.01\Omega$, $\gamma\simeq0.01\Omega$, we can have $n\leq7$ while still keeping the number of experimental runs reasonably low, since $e^{2f}\lesssim100$ (recall that a number of experimental runs $M\gg e^{2f}$ is required to filter out decoherence from the measured data). In a later section of this paper we will show how this parameter range is easily achievable for the physical system proposed in \cite{flux-oscill}. This would allow to implement our protocol for phase space points of modulus $|\xi|=nr\lesssim3.5$, as we show in Fig.~\ref{fock-plots} for the particular case of the Fock state $\ket5$.
\section{Measuring low order moments}
It is often the case that full state reconstruction is not possible and only the characteristic function in the vicinity of the origin is availabe. This might still be sufficient to evaluate low order moments of the quadrature operators $\hat X_\theta=ae^{-i\theta}+a^\dagger e^{i\theta}$. To show this, we expand the right hand side of (\ref{main-result}) in powers of $r=|\xi|$. Separating real and imaginary parts, we have a direct connection between the oscillator moments and the expectations of pauli operators:
\begin{align}
&\av{\sigma_x(t)}e^{f(g,t)}=1-\frac{1}{2}r^2\av{\hat X_\theta^2}+\frac{1}{24}r^4\av{\hat X_\theta^4}+...,\\
&\av{\sigma_y(t)}e^{f(g,t)}=-r\av{\hat X_\theta}+\frac{1}{6}r^3\av{\hat X_\theta^3}+...,
\end{align}
where $\theta=\arg\{\xi\}+\pi/2$. Low order moments of an arbitrary quadrature can therefore be obtained by first correcting the data for decoherence and subsequently performing a polynomial fit with respect to the variable $r$. The number of moments that can be reliably estimated with this method will depend on the available range of the parameter $r$. Useful information can be extracted even from the first few moments: while second order moments are sufficient to test squeezing, third and higher order moments can be used to investigate non-Gaussianity.
\section{Superposition states}
\begin{figure}[t!]
\includegraphics[width=0.24\textwidth]{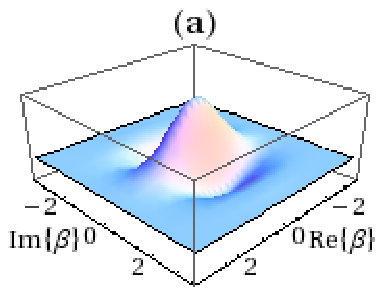}\includegraphics[width=0.24\textwidth]{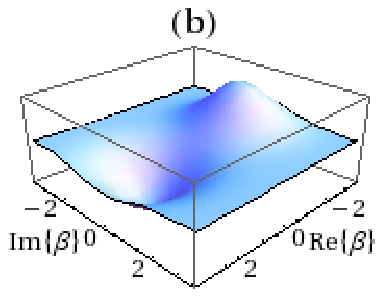}\\
\includegraphics[width=0.24\textwidth]{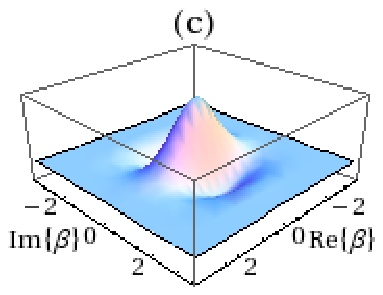}\includegraphics[width=0.24\textwidth]{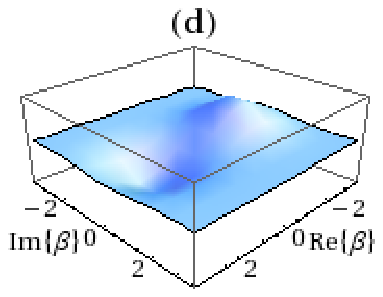}
\caption{Comparison of the characteristic functions relative to the ideal superposition state and the prepared state. Plots (a) and (b) show respectively the real and imaginary parts of $\chi$ for the superposition state $\ket{\psi^+_{\alpha,\varphi}}$ with $\alpha=1,\varphi=\frac{\pi}{2}$. In plots (c) and (d) the same quantities are shown for the state prepared in the presence of decoherence, using the harmonic coupling of Eq.(\ref{g_t}) and the parameters $r_0=0,r=0.5,n=4, \Omega=2\pi\times100$MHz, $\kappa=2\pi\times50$kHz, $\gamma=\kappa\Delta=2\pi\times1$MHz. With the choice $\varphi=\pi/2$, the in-diagonal and off-diagonal terms of the density matrix are represented separately in the real and imaginary parts of $\chi$. Plot (c) shows that the state prepared with our method reproduces well the incoherent features of the superposition state, even though a slight steepening of the central peak due to heating can be seen in comparison to plot (a). On the other hand, the interference terms are more significantly damaged by decoherence, as it can be seen by comparing the imaginary parts (b) and (d).\label{cat-fig}}
\end{figure}

One of the goals in experiments with nanomechanical oscillators is the preparation and verification of coherent superpositions of classically distinct states, in order to explore the validity of the superposition principle for macroscopic objects. In the previous sections, we fully addressed the problem of verification for arbitrary states. Here, we show that the preparation of motional superposition states can be achieved in the same framework. These possibilities together make our system a powerful toolbox for the investigation of nonclassical states of motion. We can prepare a superposition state in our system by following a few simple steps, similarly to what has been proposed in \cite{ion-trap} and \cite{Agarwal-Solano}. We suppose that the system is initialized in the state (\ref{initial}), and we assume that the oscillator has been pre-cooled to the ground state $\rho_0=\kebra{0}{0}$. At $t=0$ the cooling mechanism is switched off so that the time evolution of the system is described by (\ref{master}). We let the coupled system evolve for a time $t$, then we measure the qubit in the basis $\ket{\varphi_\pm}=\frac{1}{\sqrt2}\left(\ket g\pm e^{i\varphi}\ket e\right)$. In the absence of decoherence, the oscillator would be projected in the superposition state
\begin{equation}
\ket{\psi_{\alpha,\varphi}^\pm}\propto\ket\alpha\pm e^{-i\varphi}\ket{-\alpha},\label{ideal-cat}
\end{equation}
where $\ket{\pm\alpha}$ are coherent states with $\alpha=i\int_0^tdsg(s)e^{i\Omega s}$. The characteristic function is 
\begin{align}
&\chi_{\alpha,\varphi}^\pm(\beta)=\frac{e^{-\frac{1}{2}|\beta|^2}\cos\left(2\Im m\{\alpha\beta^*\}\right)}{1+e^{-2|\alpha|^2}\cos\varphi}+\nonumber\\
&\phantom{\chi_{\alpha,\varphi}(\beta)}\pm\frac{e^{-i\varphi-\frac{1}{2}|\beta-2\alpha|^2}+e^{i\varphi-\frac{1}{2}|\beta+2\alpha|^2}}{2+2e^{-2|\alpha|^2}\cos\varphi}\label{cat-chi}.
\end{align}
We can see that the diagonal terms $\kebra{\pm\alpha}{\pm\alpha}$ correspond to a sinusoidally modulated Gaussian peak centered in zero, while the interference terms $\kebra{\pm\alpha}{\mp\alpha}$ yield two Gaussian peaks centered at $\beta=\pm2\alpha$. The complex phase of these peaks is controlled by the relative phase $\varphi$. If we take into account finite decoherence, the described protocol yields an imperfect superposition state (related calculational details can be found in the Appendix). In particular, we can expect a steepening of the central peak due to heating (i.e. the state becomes mixed) as well as a reduction in the height of the interference peaks, due to dephasing. This is shown in Fig.~\ref{cat-fig}, where the characteristic functions obtained with our protocol in the ideal and finite-decoherence cases are compared. To conclude the present section, we point out the dual behavior of the characteristic function as compared to the Wigner function \cite{barnett}. In the latter representation, interference terms appear as oscillations, while diagonal terms yield non-centered Gaussian peaks. Moreover, in the Wigner function heating results in broadening of the Gaussian peaks rather than steepening. These observations are all consistent with the fact that the two representations are connected by a symplectic Fourier transform.
\section{Possible implementations}
Being rather general, our reconstruction method can be applied to any experimental setup in which the qubit-oscillator coupling can be coherently tuned over time. We shall give some examples of realistic settings in which such level of control can be achieved.

As our main example, we consider the theoretical proposal by Fei Xue {\em et al.} in the field of nanomechanics \cite{flux-oscill}. In their paper, the coupling between a flux qubit \cite{supercond} and a nanomechanical resonator is controlled via the amplitude of an external magnetic field, according to
\begin{equation}
g(t)=\eta B(t), \label{magnetic}
\end{equation}
where $\eta$ is a constant that depends on the specific system realization, while $B(t)$ is the magnetic field amplitude along an appropriate direction. They estimate $\eta\simeq2\pi\cdot0.8$MHz$/$mT for a realistic choice of parameters, emphasizing how the control of such coupling can be achieved without interfering with the non-interacting part of the system Hamiltonian. From the form of Eq. (\ref{magnetic}), we see that both the magnitude and sign of the coupling can be controlled over time by tuning the external magnetic field. Still referring to \cite{flux-oscill}, we assume that the oscillator has a frequency $\Omega\simeq2\pi\times100$MHz, ground state spread $x_0\simeq2.6\times10^{-13}$m and quality factor $Q\simeq2\times10^4$, resulting in $\kappa\simeq2\pi\times50$kHz and $\kappa\Delta\simeq2\pi\times1$MHz at a temperature $T\sim100$mK. Assuming a maximum magnetic field intensity $B_\textrm{max}\simeq10$mT, we get $g_\textrm{max}\simeq2\pi\times8$MHz. With such parameters,  in (\ref{g_t}) we can set $r_0=0$ and $r\leq2\pi g_\textrm{max}/\Omega\simeq0.5$. The implementation of the harmonic time dependence of (\ref{g_t}) requires manipulation of the magnetic field source currents at radio-frequency, which can be achieved with modest technology. For the flux qubit we take $\Gamma_1=\Gamma_2=2\pi\times0.4$MHz and $N_q\simeq0$, yielding a total dephasing rate $\gamma=\Gamma_1/2+2\Gamma_2\simeq2\pi\times1$MHz. Inserting these parameters in Eqs. (\ref{space-point}) and (\ref{damping-approx}), considering $r_\textrm{max}\simeq0.5,n=7$, and assuming the Markovian approximation to be valid, we are able to reach phase space points up to a distance $|\xi|\simeq3.5$ with a relatively low number of experimental runs, since $e^{2f}\lesssim100$ as shown in Fig.~\ref{fock-plots}. This parameter range is already sufficient to identify a large variety of states, including Fock states of low order, coherent states and superposition states with $|\alpha|\lesssim1$. To measure states with a thermal excitation, even smaller values of $|\xi|$ could be sufficient, since finite temperature induces localization of the characteristic function around the origin. As the number of mechanical periods is increased, further regions of phase space become available, however the number of required runs blows up rapidly for $n>7$, and already for $n=10$ we have $e^{2f}\sim10^5$. With the above parameters it is also possible to prepare superposition states with $|\alpha|\simeq1$, as shown in Fig.~\ref{cat-fig}.

As a second example we consider the recent experimental work of Srinivasan \textit{et al.} \cite{srinivasan}. In it, the authors demonstrate a novel circuit-QED architecture, in which both the internal levels splitting of a qubit and its dipolar coupling to a microwave cavity mode can be independently tuned over a wide range of parameters. The system was theoretically proposed in \cite{gambetta-houck}. By controlling external bias voltages, the authors are able to continuously vary $g$ in the range $\sim2\pi\cdot200$kHz$\div46$MHz. The harmonic oscillator is in this case a microwave cavity mode of frequency $\Omega\sim2\pi\cdot5$GHz, so that the ratio $g/\Omega$ varies between a negligible value and $\sim0.01$. Recent results from the same group suggest the possibility to reach $g\sim2\pi\cdot300$MHz, which would push the ratio up to $\sim0.06$. Even though the experiment by Srinivasan focuses on the Jaynes-Cummings regime, we argue that there is no fundamental reason why the same ideas should not work in the dispersive regime required by our protocol, and we hope to see experimental confirmations in this direction in the near future.

The application of our scheme to systems based on charge qubits \cite{LaHaye} is more problematic at the actual state of technology, essentially due to fast qubit dephasing. Moreover, the background charge noise acting on the qubit is usually not well described by a Markovian master equation of the form (\ref{master}). However, the situation might be improved by combining our protocol with charge-echo techniques \cite{charge-echo}, which we leave for future investigations.
\section{Conclusions}
Before concluding, we emphasize that estimating decoherence through Eq. (\ref{master}) is in general an approximation, since many environments are not exactly Markovian. For environments that show non-negligible deviations from Markovianity, we can expect the predictions of our model to get progressively worse as the interaction time is increased. To correct this, non Markovian effects could be included numerically in the model. 

Even when decoherence is accurately modeled by the master equation (\ref{master}), in a real experiment unavoidable errors will arise due to random fluctuations in the coupling strength $g$ and a limited accuracy in the control of the interaction time $t$. Similarly, the decoherence parameters appearing in the master equation might be known with a non-negligible uncertainty. However, such errors can be bounded, due to the continuity of the functionals in Eqs. (\ref{xi_t}-\ref{mu_t}) with respect to $(g,t)$ and the decoherence parameters.

To summarize, we presented a scheme in which a parametrically coupled qubit can be used to measure the characteristic
function of a nanomechanical oscillator. By introducing the possibility of a time-varying coupling,
we have shown how the characteristc function can be measured just by exploiting the system's time evolution, while the effect of Markovian decoherence can be filtered out by increasing the number of experimental runs.
\section{Acknowledgements}
We thank A. Ferraro and G. J. Milburn for the useful discussions, the Engineering and Physical Sciences Research Council in the United Kingdom, the
Quantum Information Processing Interdisciplinary Research Collaboration, the Royal Society and the Wolfson Foundation.
\section{Appendix: solution methods}
We solve the master equation (\ref{master}) by using phase space methods similar to those used in \cite{tufa}. We consider a representation in which a matricial characteristic function is used to describe the state of the coupled system. We decompose the total density matrix at time $t$ as
\begin{align}
&\rho_\textrm{tot}(t)=\rho_e(t)\otimes\kebra{e}{e}+\rho_g(t)\otimes\kebra{g}{g}+\nonumber\\
&\phantom{\rho_\textrm{tot}(t)}+\rho_+(t)\otimes\kebra{e}{g}+\rho_-(t)\otimes\kebra{g}{e}.
\end{align}
By defining the characteristic function for each element as $\chi_j(\beta,t)=\tra{\rho_j(t)D(\beta)}$, we can define the matricial characteristic function as:
\begin{align}
&\chi_\textrm{tot}(\beta,t)=\chi_e(\beta,t)\kebra{e}{e}+\chi_g(\beta,t)\kebra{g}{g}+\nonumber\\
&\phantom{\chi_\textrm{tot}(\beta,t)}+\chi_+(\beta,t)\kebra{e}{g}+\chi_-(\beta,t)\kebra{g}{e}.
\end{align}
At this point, we have to convert (\ref{master}) to a system of coupled partial differential equations for the functions $\chi_j$, which can be done by using standard techniques \cite{barnett}. 

The expectation values required for our state reconstruction protocol are:
\begin{align}
&\av{\sigma_x(t)}=\tr{\rho_\textrm{tot}(t)\sigma_x}=\chi_+(0,t)+\chi_-(0,t),\label{resuproof}\\
&\av{\sigma_y(t)}=\tr{\rho_\textrm{tot}(t)\sigma_y}=i\left[\chi_+(0,t)-\chi_-(0,t)\right],
\end{align}
therefore for the time being we only need to compute the evolution of the off diagonal elements $\chi_\pm(\beta,t)$. The equations for $\chi_\pm$ are already in diagonal form:
\begin{align}
&\partial_t\chi_\pm=\pm2ig(t)(e^{-i\Omega t}\partial_{\beta^*}-e^{i\Omega t}\partial_\beta)\chi_\pm+\mathcal L_m\chi_\pm-\gamma\chi_\pm,\label{chi-}
\end{align}
where the differential form of the mechanical Lindblad operator is $\mathcal L_m\chi_j=-\frac{\kappa}{2}\left(\beta\partial_\beta+\beta^*\partial_{\beta^*}+2\Delta|\beta|^2\right)\chi_j.$
The solution of Eq. (\ref{chi-}) is
\begin{align}
&\chi_\pm(\beta,t)=\chi_\pm(\beta e^{-\frac{\kappa}{2}t}\mp\xi,0)e^{-\Delta\left(1-e^{-\kappa t}\right)|\beta\mp\mu|^2-\nu},\label{sol_3}
\end{align}
where $\nu=\gamma t+\kappa\Delta\int_0^tds|\mu(g,s)|^2$, and $\xi,\mu$ are the functionals defined in (\ref{xi_t}) and (\ref{mu_t}). Now, Eq. (\ref{main-result}) follows by considering the initial conditions corresponding to the initial state (\ref{initial}), i.e. $\chi_+(\beta,0)=\chi_-(\beta,0)=\frac{1}{2}\chi(\beta)$. Note that the factorization of the effect of decoherence in the right hand side of Eq. (\ref{main-result}) is in strong analogy with the beam-splitter model for the decoherence of a bosonic mode \cite{kim-2}.

To treat the superposition state preparation, we begin by calculating the characteristic function at time $t$ after postselection of the outcome $\ket{\varphi_\pm}$
\begin{align}
&\bra{\varphi_\pm}\chi_\textrm{tot}(\beta,t)\ket{\varphi_\pm}=\chi_e(\beta,t)+\chi_g(\beta,t)+\nonumber\\
&\phantom{\bra{\varphi_+}\chi_\textrm{tot}(\beta,t)\ket{\varphi_+}}\pm e^{-i\varphi}\chi_+(\beta,t)+\pm e^{i\varphi}\chi_-(\beta,t).\label{projected}
\end{align}
After normalization, we get the characteristic function of the imperfect superposition state
\begin{equation}
 \chi_\pm=\frac{\chi_e(\beta,t)+\chi_g(\beta,t)\pm e^{-i\varphi}\chi_+(\beta,t)\pm e^{i\varphi}\chi_-(\beta,t)}{2\pm e^{-i\varphi}\chi_+(0,t)\pm e^{i\varphi}\chi_-(0,t)}.
\end{equation}
We see from the above equations that we need to compute the evolution of the remaining elements $\chi_{e,g}$. For simplicity, we neglect thermal excitations in the qubit bath and take $N_q\simeq0$. This is justified by the fact that the qubit frequency can be several orders higher than the oscillator's. With this approximation, the equations of motion are
\begin{align}
&\partial_t\chi_g=-ig(t)(e^{-i\Omega t}\beta+e^{i\Omega t}\beta^*)\chi_g+\mathcal L_m\chi_g+\Gamma_1\chi_e,\label{non-homo}\\
&\partial_t\chi_e=ig(t)(e^{-i\Omega t}\beta+e^{i\Omega t}\beta^*)\chi_e+\mathcal L_m\chi_e-\Gamma_1\chi_e\label{homo}.
\end{align}
Eq. (\ref{homo}) is homogeneous and can be readily solved
\begin{align}
&\chi_e(\beta,t)=e^{-\Gamma_1 t-\Delta\left(1-e^{-\kappa t}\right)|\beta|^2+\lambda\beta^*-\lambda^*\beta}\chi_e(\beta e^{-\frac{\kappa}{2}t},0),
\end{align}
where $\lambda=ie^{-\frac{\kappa}{2} t}\int_0^tdsg(s)e^{i\Omega s+\frac{\kappa}{2} s}$ (note that this is also a functional depending on $g$ and $t$). Eq. (\ref{non-homo}) is non homogeneous due to the presence of the term $\Gamma_1 \chi_e$, therefore yielding a more involved solution
\begin{align}
&\chi_g(\beta,t)=\bar\chi_g(\beta,t)\left[1+\Gamma_1\int_0^tds\frac{\chi_e(\beta e^{\frac{\kappa}{2}(s-t)},s)}{\bar\chi_g(\beta e^{\frac{\kappa}{2}(s-t)},s)}\right],
\end{align}
where $\bar \chi_g$ is the homogeneous solution of Eq. (\ref{non-homo}),
\begin{align}
&\bar\chi_g(\beta,t)=e^{-\Delta\left(1-e^{-\kappa t}\right)|\beta|^2+\lambda^*\beta-\lambda\beta^*}\chi_g(\beta e^{-\frac{\kappa}{2}t},0).
\end{align}
Since we supposed that the oscillator is initialized in the ground state, the initial conditions are now given by $\chi_j(\beta,0)=\frac{1}{2}e^{-\frac{1}{2}|\beta|^2}$ for $j=e,g,+,-$.


\begin{thebibliography}{99}
\bibitem{pure-qubit} The protocol can be easily generalized for non-pure states of the qubit. If the qubit is in a low temperature thermal state $w_p=p\kebra{e}{e}+(1-p)\kebra{g}{g}$, we can consider an initial qubit state $w=\mathcal Hw_p\mathcal H$, where $\mathcal H$ is the Hadamard gate given by $\mathcal H\ket g=\ket+, \mathcal H\ket e=\ket-$. As a result, the right hand side of Eq.(\ref{main-result}) gets multiplied by a factor $(1-2p)$, as it can be seen by applying the calculational methods described in the Appendix. As a further generalization, one could consider imperfect Hadamard gates, which can be treated similarly.
\bibitem{s-parametrized} If we introduce a one-to-one mapping $\xi\leftrightarrow(g_\xi,t_\xi)$, so that $f$ becomes a function of the phase space point $\xi$, the right hand side of Eq.(\ref{main-result}) can be seen as a generalized characteristic function, given by $\chi_f(\xi)=\tr{\rho_0D(\xi)}e^{-f(\xi)}$. If the function $f$ is not singular at finite $\xi$, $\chi_f(\xi)$ is clearly a valid representation of the quantum state $\rho_0$. The relation between the derivatives of $\chi_f(\xi)$ and the quadrature moments will depend upon the specific form of $f$. In the special case $f(\xi)=-s|\xi|^2/2$, we get the s-parametrized characteristic function \cite{barnett}, whose derivatives yield s-ordered quadrature moments for $-1\leq s \leq 1$.
\bibitem{explain} To prove this, we consider the difference $f\left(g_{r,\phi},t_{n+1}\right)-f\left(g_{r,\phi},t_n\right)$. The increment proportional to $\gamma$ is clearly positive, while the remaining part is given by a quadratic form in $r,r_0$. A numerical software can be used to show that this quadratic form is positive definite for any value of $\phi$ and any integer $n\geq1$.
\end{thebibliography}
\end{document}